\documentclass[aps,prb,showpacs,twocolumn]{revtex4-1}

\usepackage{ulem}
\usepackage{graphicx}
\usepackage{CJK}
\usepackage{amssymb}
\usepackage{epsfig}
\usepackage{graphicx}
\usepackage{subfigure}
\usepackage{mathrsfs}
\usepackage{verbatim}
\usepackage{rotating}
\usepackage{longtable}
\usepackage{amsmath}
\usepackage{array,color}
\usepackage{dcolumn}
\usepackage{bm}

\begin{document}
\title{Comment to the paper J. Yan et al. Experimental confirmation of driving pressure boosting and smoothing for hybrid-drive inertial fusion at the 100-kJ laser facility [Nature Communications (2023) 14:5782]}
\begin{CJK*}{GB}{gbsn}

\author{Ke Lan (À¶¿É)}\email{lan$\_$ke@iapcm.ac.cn}
\affiliation{Institute of Applied Physics and Computational Mathematics, Beijing 100094, China}


\maketitle


The authors of Ref. \citenum{Yan2023NC} reported that the driving pressure boosting and smoothing of hybrid-drive inertial fusion (HD) had been confirmed at the 100-kJ laser facility  by using the hemispherical and planar ablator targets installed in a semicylindrical hohlraum. However, in this comment we show that, due to the lack of key data on the pressure of  the direct-drive (DD) laser  under the same conditions and the lack of key data on the observed asymmetries, the  logic of this paper is not closed, so its conclusion is not  credible.

According to the authors,  in the 3rd paragraph on Page 2 of Ref. \citenum{Yan2023NC}, the HD scheme
is a coupling of  indirect-drive (ID) and DD,  and HD has two phases.
In the 1st phase, only the ID  laser beams are activated.
In the 2nd phase, the main-pulse laser of ID works together with the  DD  laser which  enters into hohlraum and irradiates directly the  capsule.
According to the authors, in the 1st paragraph of the right-hand column on Page 2 of Ref. \citenum{Yan2023NC}, the DD laser   plays an important role in the HD scheme, because it can provide a stable support for the key effect ``bulldozer''  for pressure boosting and smoothing in the second phase of HD.
According to the authors, in the last paragraph  on Page 3 of Ref. \citenum{Yan2023NC},
the HD pressure boosting and smoothing effects are at the heart of the HD scheme.

In the experiment, the authors used 43 - 50 kJ for the ID laser to generate the radiations inside hohlraum with a peak temperature of $\sim$ 200 eV, and  3.6 - 4.0 kJ for the DD laser  at an intensity of $1.8 \times 10^{15}$W/cm$^2$.
While, the laser energy used for ID  is much larger than that for the DD laser,
the ablation pressure created by the DD laser beams can be dominant. The
ID laser is used to heat the hohlraum wall  to generate the 200 eV radiations inside hohlraum. Since the wall has a much larger area than the target, the ID drive is less efficient than  the  DD laser drive.
According to Ref. \citenum{Yan2023NC}, the ID pressure is 45 Mbar; by applying in the second phase the DD laser with   intensity of $1.8 \times 10^{15}$W/cm$^2$, the authors measured the HD pressure of 170 - 180 Mbar for  the hemispherical target, and 150 - 155 Mbar for the planar target.
According to Ref. \citenum{Yan2023NC}, the hemispherical target has a higher pressure due to the spherical convergence effect. Here, we do not consider the spherical convergence effect and focus on the planar target.
While it is an important result that the shock pressure in  the planar target is about 3.3 - 3.4 times the ID pressure,
a comparison of ID and HD pressures does not inform us about the role of the DD laser in the pressure amplification.
The measurement of the DD  pressure  alone is not reported in the paper.
Here and hereafter in this comment, DD pressure refers to   pressure of the DD laser.
The question is: how  the DD pressure  compares to the ID and HD pressures?

To answer this question,   we consider a CH planar target and estimate the ID pressure at the radiation temperature of 200 eV and the   DD  pressure at laser intensity of $1.8 \times 10^{15}$W/cm$^2$ by using the expressions of Ref. \citenum{MTV}. Expression (7.94) of Ref. \citenum{MTV}  is for the dependence of  radiation ablation pressure   $P_{\texttt{ID}}$  on   hohlraum temperature $T_r$. Taking  the radiative reemission (denoted as albedo $\alpha$) into consideration,  we have:
\begin{eqnarray}\label{Eq1}
P_{\texttt{ID}} (\texttt{Mbar}) = 4.92 (1 - \alpha) (\frac{1 + Z}{A})^{-1/2} (\frac{T_r}{\texttt{heV}})^{7/2},
\end{eqnarray}
for a material of atomic number $Z$ and mass number $A$.  From our simulations, $\alpha$ is $\sim$ 0.35 for CH at 200 eV, then we have $P_{\texttt{ID}}$ $\sim$ 43.5 Mbar at 200 eV from Eq. (1).
Furthermore, from  expression (7.104) of Ref. \citenum{MTV} for the stationary ablation pressure for planar laser ablation,  noted as $P_{\texttt{DD}}$, we have:
\begin{eqnarray}\label{Eq2}
P_{\texttt{DD}} (\texttt{Mbar}) =  45  \times (A/Z)^{1/3} \times (I_L/\lambda)^{2/3},
\end{eqnarray}
here, $\lambda$ is the laser wavelength in $\mu$m, and $I_L$ is the laser intensity in $10^{15}$W/cm$^2$.
Thus, for the CH planar target,   $\lambda = 0.353$ $\mu$m and  $I_L = 1.8 \times 10^{15}$ W/cm$^2$, we have $P_{\texttt{DD}}$ =  164 Mbar. Considering the laser injection angle of 28.5$^\circ$ and the averaged laser backscattering fraction  of $4.5\%$ in Ref. \citenum{Yan2023NC},  we have $P_{\texttt{DD}}$ = 146 Mbar for  the CH planar target, $\sim$ 3.36 times $P_{\texttt{ID}}$.

This simple estimate of the ID pressure  agrees well with the value reported in  Ref. \citenum{Yan2023NC}. The pressure of the DD laser drive is very close to the HD pressure reported in Ref. \citenum{Yan2023NC}.
This comparison leads us to the conclusion that,  even without ID, i.e., without the first phase of HD, the pressure generated in the CH target by a DD laser  can be   3.36 times the pressure generated by ID.  Then, our question is: if  the HD pressure is  almost the same as the DD laser drive-only,  what is the benefit of HD and why the first phase of ID  is needed? Not to mention that, according to Ref. \citenum{Yan2023NC}, ID  creates an asymmetric   shock, which needs to be corrected by the DD laser drive. Then, why not use the DD laser drive-only? and what is the significance of HD?

This hypothesis of the dominance of  the pressure of DD laser  is confirmed in Fig. 3b, which is a key figure for the experimental confirmation of Ref. \citenum{Yan2023NC}. According to the authors' instructions, there are ID and HD shocks in this figure, but there is no DD shock at all.

In addition, the authors of Ref. \citenum{Yan2023NC} claim  that their experimental results
demonstrate that the HD scheme  can provide a smoothed   pressure compared to the radiation ablation
pressure, but they did not present any  experimental evidence.
In their description on the pressure smoothing of the HD scheme, the authors only mention the  pressure  asymmetries   of HD and ID, but not that of the DD laser pulse. This looks like the pressure  asymmetry of the DD laser can be completely ignored in the HD scheme. This is in odds with the DD laser implosion, which creates stronger asymmetries because of a limited number of laser beams,  the crossed-beam energy transfer  and the laser imprint.

The authors claim that the HD scheme is a coupling of the ID and the DD laser drive.
Would it be so, the HD pressure should be superior to the sum of ID   and DD pressures.
Otherwise, why not use  the total laser energy of HD for the DD laser drive-only or for the ID-only?
Hence, to verify the advantages of HD in pressure boosting, the authors should measure all the three  pressures of ID, HD and DD.
The same argument applies  to the pressure smoothing, i.e.,  the experimental evidence of the asymmetries of ID, HD and   DD  should be presented for   confirmation.

The lack of key data on the pressure of  the DD laser  under the same conditions and the lack of key data on the   asymmetries  lead us to the conclusion that, the published experimental confirmation of driving pressure boosting and smoothing for hybrid-drive inertial fusion is not credible.

In the last paragraph above {\bf Methods} on Page 7 of Ref. \citenum{Yan2023NC}, the authors write that ``We found from experimental data and the simulation results of the ignition targets that there is an approximately fitted hydroscaling relationship in the form $P_{\texttt{HD}}$ (Mbar) $\approx$  180 $(E_{\texttt{DD}}/4)^{1/4}$ $(T_r/2)$'', where $E_{\texttt{DD}}$ is the DD laser energy, and the authors emphasize in their response to the referee reports \cite{response} that the HD pressure is independent of the laser intensity.
No matter how the  scaling relationship is obtained, it must not violate the basic  physics.  For the same target and under the same   DD laser energy, how can the HD pressure remain the same if one changes the laser pulse duration?

{\bf ACKNOWLEDGMENTS}
This work is supported by the National Natural Science Foundation of China (Grant No. 12035002).
\end{CJK*}

\end{document}